\documentclass[prc,twocolumn,aps,superscriptaddress,showpacs]{revtex4}
\usepackage{amssymb}
\usepackage{amsmath,bm}
\usepackage{graphicx}
\usepackage[normalem]{ulem}
\usepackage{booktabs}
\usepackage{threeparttable}
\usepackage[dvips]{color}
\usepackage{leftidx}
\renewcommand\sout{\bgroup \color{red} \ULdepth=-.5ex \ULset}

\begin{document}

\title{Nuclear matter fourth-order symmetry energy in non-relativistic mean-field models}
\author{Jie Pu}
\affiliation{School of Physics and Astronomy and Shanghai Key Laboratory for
Particle Physics and Cosmology, Shanghai Jiao Tong University, Shanghai 200240, China}
\author{Zhen Zhang}
\affiliation{Cyclotron Institute and Department of Physics and Astronomy, Texas A\&M University, College Station, Texas 77843, USA}
\author{Lie-Wen Chen\footnote{%
Corresponding author (email: lwchen$@$sjtu.edu.cn)}}
\affiliation{School of Physics and Astronomy and Shanghai Key Laboratory for
Particle Physics and Cosmology, Shanghai Jiao Tong University, Shanghai 200240, China}
\affiliation{Center of Theoretical Nuclear Physics, National Laboratory of Heavy Ion
Accelerator, Lanzhou 730000, China}
\date{\today}

\begin{abstract}
\noindent
\textbf{Background:} Nuclear matter fourth-order symmetry energy $E_{\text{sym,4}}(\rho)$ may significantly influence the properties of neutron stars such as the core-crust transition density and pressure as well as the proton fraction at high densities. The magnitude of $E_{\text{sym,4}}(\rho)$ is, however, largely uncertain.

\noindent
\textbf{Purpose:} Based on systematic analyses of several popular non-relativistic energy density functionals with mean-field approximation, we estimate the value of the $E_{\text{sym,4}}(\rho)$ at nuclear normal density $\rho_0$ and its density dependence, and explore the correlation between $E_{\text{sym,4}}(\rho_0)$ and other macroscopic quantities of nuclear matter properties.

\noindent
\textbf{Method:} We use the empirical values of some nuclear macroscopic quantities to construct model parameter sets by  Monte Carlo method for four different energy density functionals with mean-field approximation, namely, the conventional Skyrme-Hartree-Fock (SHF) model, the extended Skyrme-Hartree-Fock (eSHF) model, the Gogny-Hartree-Fock (GHF) model, and the momentum-dependent interaction (MDI) model. With the constructed samples of parameter sets, we can estimate the density dependence of $E_{\text{sym,4}}(\rho)$ and analyze the correlation of $E_{\text{sym,4}}(\rho_0)$ with other macroscopic quantities.

\noindent
\textbf{Results:} The value of $E_{\text{sym,4}}(\rho_0)$ is estimated to be $1.02\pm0.49$ MeV for the SHF model, $1.02\pm0.50$ MeV for the eSHF model, $0.70\pm0.60$ MeV for the GHF model, and $0.74\pm0.63$ MeV for the MDI model.
Moreover, our results indicate that the density dependence of $E_{\text{sym,4}}(\rho)$ is model dependent, especially at higher densities. Furthermore, we find that the $E_{\text{sym},4}(\rho_0)$ has strong positive (negative) correlation with isoscalar (isovector) nucleon effective mass $m_{s,0}^*$ ($m_{v,0}^*$) at $\rho_0$. In particular, for the SHF and eSHF models, the $E_{\text{sym,4}}(\rho)$ is completely determined by the isoscalar and isovector nucleon effective masses $m_{s}^*(\rho)$ and $m_{v}^*(\rho)$, and the analytical expression is given.

\noindent
\textbf{Conclusions:} In the mean-field models, the magnitude of $ E_{\text{sym,4}}(\rho_0)$ is generally less than $2$ MeV, and its density dependence depends on models, especially at higher densities. $ E_{\text{sym,4}}(\rho_0)$ is strongly correlated with $m_{s,0}^*$ and $m_{v,0}^*$.

\end{abstract}

\maketitle

\section{Introduction}

The equation of state (EOS) of isospin asymmetric nuclear matter $E(\rho,\delta)$, defined as the binding energy per nucleon, is one of fundamental issues in both nuclear physics and astrophysics~\cite{LiBA98,Dan02,Lat04,Ste05,Bar05,LCK08,Tra12,Hor14,LiBA14,Heb15,Bal16,Oer17,LiBA17}.
In the widely-used parabolic approximation, namely, $E(\rho,\delta) \approx E_0(\rho)+E_{\mathrm{sym}}(\rho)\delta^2$, where $\rho$ is the total nucleon density and $\delta=(\rho_n-\rho_p)/\rho$ is the isospin asymmetry with $\rho_{n(p)}$ being the neutron (proton) density, the nuclear matter symmetry energy $E_{\mathrm{sym}}(\rho)$ determines the isospin dependence of nuclear matter EOS. During the past decades, a lot of work has been devoted to exploring the density dependence of the $E_{\text{sym}}(\rho)$ from various aspects including theory, experiment and astrophysical observation, and significant progress has been made (see, e.g., Ref.~\cite{LiBA14}).
While the parabolic approximation has been shown to be very successful, it breaks down in some special cases. For example, in the study of neutron stars where the isospin asymmetry $\delta$ could be close to unity, the higher-order terms in $\delta$ presented in the EOS of asymmetric nuclear matter, e.g., the fourth-order term $E_{\mathrm{sym,4}}(\rho)\delta^4$ with the $E_{\mathrm{sym,4}}(\rho)$ denoted as the fourth-order symmetry energy, may significantly affect the core-crust transition density and pressure, the proton fraction in $\beta$-equilibrium neutron star matter, and the critical density for the direct Urca process which can lead to faster cooling of neutron stars~\cite{Zha01,Ste06,XuJ09,Cai12,Sei14,Gon17}. As clearly demonstrated in Ref.~\cite{XuJ09}, the reason why the core-crust transition density and pressure are sensitive to the higher-order symmetry energies (although they are quite small compared to the symmetry energy $E_{\mathrm{sym}}(\rho)$) is mainly due to the fact that these issues are related to the first- and second-order derivatives of the energy with respect to the isospin asymmetry $\delta$. In addition, the higher-order symmetry energies may have comparable magnitude compared with the $E_{\mathrm{sym}}(\rho)$ at higher densities, especially in the case for a softer symmetry energy $E_{\mathrm{sym}}(\rho)$, and thus they may influence the proton fraction in $\beta$-equilibrium neutron star matter at higher densities (see, e.g., Refs.~\cite{Zha01,Cai12}).

However, so far there are essentially no empirical information on the higher-order nuclear matter symmetry energies. For instance, some recent studies predict quite different values of the fourth-order symmetry energy $E_{\mathrm{sym,4}}(\rho_0)$ at nuclear normal (saturation) density $\rho_0$.  While calculations using the non-relativistic mean-field model~\cite{Che09,Agr17}, the relativistic mean-field model~\cite{Cai12}
and the chiral pion-nucleon dynamics~\cite{Kai15} indicate that
the $E_{\mathrm{sym,4}}(\rho_0)$  is less than 2 MeV,
a study  within the quantum molecular dynamics predicts
$E_{\mathrm{sym,4}}(\rho_0)=3.27\sim12.7$ MeV depending
on the interaction used \cite{Nan16}.
In Ref.~\cite{Cai15}, the kinetic part of $E_{\mathrm{sym,4}}(\rho_0)$ is predicted to be $7.18\pm2.52$ MeV by considering the high-momentum tail in the single-nucleon momentum distributions based on an interacting Fermi gas model that could be due to short-range correlations of nucleon-nucleon interactions.
Most recently, a significantly large value of $E_{\text{sym,4}}(\rho_0)= 20.0\pm4.6 $ is estimated within an extended semi-empirical nuclear mass formula~\cite{Wan17} by analyzing the fourth-order symmetry energy of finite nuclei~\cite{Jia14,Jia15,Wan15,Tia16} extracted from nuclear mass data.
Given such a large uncertainty, a systematic study on the fourth-order symmetry energy is therefore critically important, and this provides the main motivation of the present work.

In this work, we employ four energy density functionals within non-relativistic mean-field models to study the value of the fourth-order symmetry energy at nuclear normal density as well as its density dependence.
We find in the four mean-field models the value of $ E_{\text{sym,4}}(\rho_0)$ is generally less than $2$ MeV, but the density dependence of $ E_{\text{sym,4}}(\rho)$ is model dependent.
The correlations of  the $E_{\text{sym,4}}(\rho_0)$ to other macroscopic quantities are also examined, and we find the $E_{\text{sym,4}}(\rho_0)$ is strongly correlated with the isocalar and isovector nucleon effective masses.

The paper is organized as follows.
Section~\ref{Sec2} introduces the four energy density functionals within mean-field models, i.e., the Skyrme-Hartree-Fock (SHF) model, the extended Skyrme-Hartree-Fock (eSHF) model, the Gogny-Hartree-Fock (GHF) model and the momentum-dependent interaction (MDI) model, and gives the corresponding explicit expressions of the symmetry energy and the fourth-order symmetry energy.
In Section~\ref{Sec3}, we present results and discussions about the fourth-order symmetry energy.
Our conclusions are summarized in Section~\ref{Sec4}.

\section{Model and method}
\label{Sec2}

\subsection{Characteristic parameters of asymmetric nuclear matter}
The EOS of isospin asymmetric nuclear matter can be expanded as a power series of even-order terms in $\delta$, i.e.,
\begin{equation}\label{EOS}
\begin{split}
  E(\rho,\delta)=&E_0(\rho)+E_{\text{sym}}(\rho)\delta^2+E_{\text{sym},4}(\rho)\delta^4+{\cal O}(\delta^6),
\end{split}
\end{equation}%
where $E_0(\rho)=E(\rho,\delta=0)$ is the EOS of symmetric nuclear matter (SNM), and the symmetry energy $E_{\text{sym}}(\rho)$ and the fourth-order symmetry energy $E_{\text{sym},4}(\rho)$ are given, respectively, by
\begin{equation}\label{Esym}
\begin{split}
  E_{\text{sym}}(\rho)=\frac{1}{2!}\frac{\partial^2E\left(\rho,\delta\right)}{\partial\delta^2}\bigg|_{\delta=0},\\
\end{split}
\end{equation}%
\begin{equation}\label{Esym4}
\begin{split}
  E_{\text{sym},4}(\rho)=\frac{1}{4!}\frac{\partial^4E\left(\rho,\delta\right)}{\partial\delta^4}\bigg|_{\delta=0}.\\
\end{split}
\end{equation}%
The $E_0(\rho)$ can also be expanded, e.g., up to 3rd-order in density, around nuclear saturation density $\rho_0$ as
\begin{equation}\label{E0}
\begin{split}
  E_0(\rho)=E_0(\rho_0)+\frac{K_0}{2!}\chi^2+\frac{J_0}{3!}\chi^3+{\cal O}(\chi^4),\\
\end{split}
\end{equation}%
where $\chi=(\rho-\rho_0)/3\rho_0$ is a dimensionless variable characterizing the deviations of the density from the saturation density $\rho_0$, $E_0(\rho_0)$ is the binding energy per nucleon of SNM at $\rho_0$, and the well-known incompressibility coefficient $K_0$ and the skewness coefficient $J_0$ are given as
\begin{equation}\label{k0}
\begin{split}
  K_0=9\rho_0^2\frac{\mathrm{d}^2E_0(\rho)}{\mathrm{d}\rho_2}\bigg|_{\rho=\rho_0},\\
\end{split}
\end{equation}%
\begin{equation}\label{J0}
\begin{split}
  J_0=27\rho_0^3\frac{\mathrm{d}^3E_0(\rho)}{\mathrm{d}\rho_3}\bigg|_{\rho=\rho_0}.\\
\end{split}
\end{equation}%

Similarly, around nuclear saturation density $\rho_0$, the $E_{\text{sym}}(\rho)$ can be expanded as
\begin{equation}\label{Esymrho}
\begin{split}
E_{\text{sym}}(\rho)=&E_{\text{sym}}(\rho_0)+L \chi+\frac{K_{\text{sym}}}{2!}\chi^2 +{\cal O}(\chi^4),
\end{split}
\end{equation}%
where the $E_{\text{sym}}(\rho_0)$ is the symmetry energy at nuclear saturation density $\rho_0$, and the density slope parameter $L$ and the density curvature parameter $K_{\text{sym}}$ are defined as
\begin{equation}\label{L}
\begin{split}
  L=3\rho_0\frac{\mathrm{d}E_{\text{sym}}(\rho)}{\mathrm{d}\rho}\bigg|_{\rho=\rho_0},\\
\end{split}
\end{equation}%
\begin{equation}\label{k}
\begin{split}
  K_{\text{sym}}=9\rho_0^2\frac{\mathrm{d}^2E_{\text{sym}}(\rho)}{\mathrm{d}\rho_2}\bigg|_{\rho=\rho_0}.\\
\end{split}
\end{equation}%

The parameters $\rho_0$, $E_0(\rho_0)$, $K_0$, $J_0$, $E_{\text{sym}}(\rho_0)$, $L$, and $K_{\text{sym}}$ defined above are the lower order bulk parameters that characterize quantitatively the EOS of asymmetric nuclear matter around nuclear saturation density $\rho_0$. In particular, it has been shown in Ref.~\cite{Chen11} that the three parameters $E_0(\rho_0)$, $K_0$, and $J_0$ can reasonably characterize the EOS $E_0(\rho)$ of symmetric nuclear matter up to a density of $2\rho_0 $ while the symmetry energy $E_{\text{sym}}(\rho)$ with the density up to $2\rho_0 $ can be nicely described by the three parameters $E_{\text{sym}}(\rho_0)$, $L$, and $K_{\text{sym}}$.

\subsection{The Skyrme-Hartree-Fock model}
In the standard SHF model, nucleons generally interact with each other through the so-called standard Skyrme interaction (see, e.g., Ref.~\cite{Cha97}), i.e.,
\begin{eqnarray}\label{interactionSHF}
%\begin{split}
  V_{\text{SHF}}\left( \boldsymbol{r}_{1},\boldsymbol{r}_{2} \right)&=& t_{0}\left( 1+x_{0}P_{\sigma } \right)\delta \left( \boldsymbol{r} \right)\notag\\
  &+& \frac{1}{2}t_{1}\left( 1+x_{1}P_{\sigma } \right)\left[ \boldsymbol{k}^{'2}\delta \left( \boldsymbol{r} \right)+ \delta \left( \boldsymbol{r} \right)\boldsymbol{k}^{2} \right]\notag\\
  &+&t_{2}\left(1+x_{2}P_{\sigma } \right)\boldsymbol{k}^{'}\cdot\delta \left( \boldsymbol{r} \right)\boldsymbol{k} \notag\\
  &+&\frac{1}{6}t_{3}\left( 1+x_{3}P_{\sigma } \right)\left[\rho\left(\boldsymbol{R}\right)\right]^{\alpha}\delta \left(\boldsymbol{r} \right)\notag\\
  &+&iW_0\boldsymbol{\sigma}\cdot\left[\boldsymbol{k}^{'}\times\delta \left( \boldsymbol{r} \right)\boldsymbol{k}\right].
%\end{split}
\end{eqnarray}%
where $\boldsymbol{r}=\boldsymbol{r_1}-\boldsymbol{r_2}$, $\boldsymbol{R}=(\boldsymbol{r_1}+\boldsymbol{r_2})/2$, $P_{\sigma }$ is the spin-exchange operator, $\boldsymbol{k}=-i(\boldsymbol{\nabla_1}-\boldsymbol{\nabla_2)}/2$ is the relative momentum operator, and $\boldsymbol{k^{'}}$ is the conjugate operator of $\boldsymbol{k}$ acting on the left, $\boldsymbol{\sigma}=\boldsymbol{\sigma_1}+\boldsymbol{\sigma_2}$ is the Pauli spin operator, $t_0\sim t_3$, $x_0\sim x_3$ and $\alpha$ are Skyrme force parameters, and $W_0$ is the spin-orbit coupling constant.
The EOS of asymmetric nuclear matter can be written as
\begin{equation}\label{EOSSHF}
\begin{split}
E(\rho,\delta)&= \frac{3\hbar^{2}}{10m}k_{F}^{2}F_{5/3}\\
& + \frac{1}{8}t_{0}\rho\left[2\left(x_{0}+2\right)-\left(2x_{0}+1\right)F_{2}\right]\\
& +\frac{1}{48}t_{3}\rho^{\alpha+1}\left[2\left(x_{3}+2\right)-\left(2x_{3}+1\right)F_{2}\right]\\
& +\frac{3}{40}\rho k_{F}^{2}\left[t_{1}\left(x_{1}+2\right)+t_{2}\left(x_{2}+2\right)\right]F_{5/3}\\
& +\frac{3}{80}\rho k_{F}^{2}\left[t_{2}\left(2x_{2}+1\right)-t_{1}\left(2x_{1}+1\right)\right]F_{8/3} ,
\end{split}
\end{equation}%
where $m$ is the nucleon mass, $k_{F}=\left(\frac{3\pi^{2}}{2} \rho \right)^{1/3}$ is the Fermi momentum of symmetric nuclear matter,
and $ F_{x}\left(\delta\right) $ is expressed as
\begin{equation}\label{Fx}
F_{x}\left(\delta\right)=\frac{1}{2}\left[\left(1+\delta\right)^{x}+\left(1-\delta\right)^{x}\right].
\end{equation}%
In the SHF model, the nuclear symmetry energy is given by
\begin{equation}\label{SHFEsym}
\begin{split}
E_{\text{sym}}(\rho)&=\frac{\hbar^{2}}{6m} k_{F}^{2}-\frac{1}{8} t_{0}\left(2x_{0}+1\right)\rho\\
  &-\frac{1}{48}t_{3}(2x_{3}+1)\rho^{\alpha+1}\\
  &-\frac{1}{24}\left[3t_{1}x_{1}-t_{2}\left(4+5x_{2}\right)\right]\rho k_{F}^{2} ,
\end{split}
\end{equation}
and the fourth-order symmetry energy is expressed as~\cite{Che09} (Note: There are typos in the $E_{\text{sym,4}}(\rho)$ expression in  Ref.~\cite{Che09}, namely, the `1' and `2' in $\Theta_{\rm sym,4}$ defined there should be exchanged.)
\begin{equation}\label{SHFEsym4}
\begin{split}
  E_{\text{sym,4}}(\rho)&=\frac{\hbar^{2}}{162m}k_{F}^{2}\\
  &+\frac{1}{648}k_{F}^{2}\left[3t_{1}\left(1+x_{1}\right)+t_{2}\left(1-x_{2}\right)\right]\rho .
\end{split}
\end{equation}%

\subsection{The extended Skyrme-Hartree-Fock model}
With the inclusion of additional zero-range density- and momentum-dependent terms which effectively simulate the momentum-dependent three-body force~\cite{Kre77,Ge86,Zhu88,Cha09,Gor10,Gor13,Gor15,Zha16},
the extended Skyrme interaction in the eSHF model has the following form
\begin{equation}\label{interactioneSHF}
\begin{split}
  V_{\text{eSHF}}\left( \boldsymbol{r}_{1},\boldsymbol{r}_{2} \right)&= V_{\text{SHF}}\left( \boldsymbol{r}_{1},\boldsymbol{r}_{2} \right) +\frac{1}{2}t_{4}\left(1+x_{4}P_{\sigma}\right)\\
  &\times\left[\boldsymbol{k^{'2}}\rho^{\beta}\left(\frac{\boldsymbol{r}_{1}+\boldsymbol{r}_{2}}{2}\right)\delta\left(\boldsymbol{r}\right)+\delta \left( \boldsymbol{r} \right)\boldsymbol{k}^{2}\right]\\
  &+t_{5} \left(1+x_{5}P_{\sigma}\right)\boldsymbol{k^{'}}\cdot\rho^{\gamma}\left(\frac{\boldsymbol{r}_{1}+\boldsymbol{r}_{2}}{2}\right)\delta\left(\boldsymbol{r}\right)\boldsymbol{k}.
\end{split}
\end{equation}
Therefore, the present extended Skyrme interaction has  thirteen Skyrme parameters, i.e., $t_0\sim t_5$, $x_0\sim x_5$, $\alpha$, $\beta$ and $\gamma$.
The EOS of asymmetric nuclear matter can be written as
\begin{equation}\label{eshfEOS}
\begin{split}
E(\rho,\delta)&=\frac{3\hbar^{2}}{10m}k_{F}^{2}F_{5/3}\\
& + \frac{1}{8}t_{0}\rho\left[2\left(x_{0}+2\right)-\left(2x_{0}+1\right)F_{2}\right]\\
& +\frac{1}{48}t_{3}\rho^{\alpha+1}\left[2\left(x_{3}+2\right)-\left(2x_{3}+1\right)F_{2}\right]\\
& +\frac{3}{40}\rho k_{F}^{2}\left[t_{1}\left(x_{1}+2\right)+t_{2}\left(x_{2}+2\right)\right]F_{5/3}\\
& +\frac{3}{80}\rho k_{F}^{2}\left[t_{2}\left(2x_{2}+1\right)-t_{1}\left(2x_{1}+1\right)\right]F_{8/3}\\
 &+\frac{3}{40}\rho k_{F}^{2} \left[t_{4}\left( x_{4}+2 \right) \rho^{\beta} + t_{5}\left( x_{5}+2 \right)\rho^{\gamma}\right]F_{5/3}\\
 & +\frac{3}{80}\rho k_{F}^{2} \left[ t_{5}\left( 2x_{5}+1\right) \rho^{\gamma} - t_{4}\left( 2x_{4}+1 \right) \rho^{\beta} \right] F_{8/3} ,
\end{split}
\end{equation}
In the eSHF model, the symmetry enery is expressed as
\begin{equation}\label{eshfEsym}
\begin{split}
E_{\text{sym}}(\rho)&=\frac{\hbar^{2}}{6m} k_{F}^{2}-\frac{1}{8} t_{0}\left(2x_{0}+1\right)\rho\\
  &-\frac{1}{48}t_{3}(2x_{3}+1)\rho^{\alpha+1}\\
  &-\frac{1}{24}\left[3t_{1}x_{1}-t_{2}\left(4+5x_{2}\right)\right]\rho k_{F}^{2}\\
&-\frac{1}{24}\left[3t_{4}x_{4}\rho^{\beta}-t_{5}\left(4+5x_{5}\right)\rho^{\gamma}\right]\rho k_{F}^{2} ,
\end{split}
\end{equation}
and the fourth-order symmetry energy is expressed as
\begin{equation}\label{eshfEsym4}
\begin{split}
E_{\text{sym,4}}(\rho)&=\frac{\hbar^{2}}{162m}k_{F}^{2}\\
  &+\frac{1}{648}k_{F}^{2}\left[3t_{1}\left(1+x_{1}\right)+t_{2}\left(1-x_{2}\right)\right]\rho\\
&+\frac{1}{648}\cdot 3t_{4}\left(1+x_{4}\right)k_{F}^{2}\rho^{1+\beta}\\
&+\frac{1}{648}t_{5}\left(1-x_{5}\right)k_{F}^{2}\rho^{1+\gamma}.
\end{split}
\end{equation}

\subsection{The Gogny-Hartree-Fock model}
In the GHF model, the Gogny interaction~\cite{Dec80} is expressed as
\begin{eqnarray}\label{gogny}
% \begin{split}
 V_{12}(\boldsymbol{r})&=&\sum_{i=1,2}\left(W_i+B_i P_\sigma-H_i P_\tau-M_i P_\sigma P_\tau \right)e^{-\frac{\boldsymbol{r}^2}{\mu_i^2}} \notag\\
 &+&t_0\left(1+x_0 P_\sigma\right)\rho^\alpha \left( \vec{R} \right) \delta \left( \vec{r_1}-\vec{r_2}\right) \notag\\
 &+&iW_0\boldsymbol{\sigma}\cdot\left[\boldsymbol{k}^{'}\times\delta \left( \boldsymbol{r} \right)\boldsymbol{k}\right],
% \end{split}
\end{eqnarray}
where $W_1$, $B_1$, $H_1$, $M_1$, $\mu_1$, $W_2$, $B_2$, $H_2$, $M_2$, $\mu_2$, $t_0$, $x_0$ and $\alpha$ are 13 Gogny interaction parameters, and $P_\tau$ is the isospin exchange operator. The Gogny interaction has a zero-range density dependent term which is helpful in reproducing saturation properties of nuclear matter, together with a finite-range part modeled by two Gaussian functions that can simulate the short- and middle-range parts of nuclear forces.
By using the Hartree-Fork approach, the EOS of asymmetric nuclear matter with Gogny interaction is expressed as~\cite{Che12,Sel14}
\begin{widetext}
\begin{equation}\label{gognyEOS}
\begin{split}
&E\left(\rho,\delta\right)=\frac{3}{5}\frac{\hbar^{2}k_{F}^{2}}{2m}\frac{1}{2}\left[\left(1+\delta\right)^{5/3}+\left(1-\delta\right)^{5/3}\right]\\
&+\frac{1}{2}\sum_{i=1,2}\left\lbrace \left[ \frac{\pi^{3/2}\mu_{i}^{3}}{4}\left(4W_i+2B_i-2H_i-M_i\right)+\frac{3}{4}t_0 \right]\rho+\left[\frac{\pi^{3/2}\mu_i^3}{4}\left(-2H_i-M_i\right)-\frac{1}{4}t_0\left(1+2x_0\right)\rho^{\alpha_i}\right]\rho\delta^2\right\rbrace\\
&+\frac{1}{2}\sum_{i=1,2}\left\lbrace-\frac{1}{\sqrt{\pi}}\left(W_i+2B_i-H_i-2M_i\right)\right.\cdot\left[\frac{1+\delta}{2}\Bigg(\frac{2}{\left(\mu_i k_F^n\right)^3}-\frac{3}{\mu_i k_F^n}-\frac{2}{\left(\mu_i k_F^n\right)^3}e^{-q^2}\right.+\frac{1}{\mu_i k_F^n}e^{-q^2}+\sqrt{\pi}\text{erf}\left(\mu_i k_F^n\right)\Bigg)\\
&+\frac{1-\delta}{2}\Bigg(\frac{2}{\left(\mu_i k_F^p\right)^3}-\frac{3}{\mu_i k_F^p}-\frac{2}{\left(\mu_i k_F^p\right)^3}e^{-q^2}\left.+\frac{1}{\mu_i k_F^p}e^{-q^2}+\sqrt{\pi}\text{erf}\left(\mu_i k_F^p\right)\Bigg)\right]\left.+\frac{1}{\sqrt{\pi}}\left(H_i+2M_i\right)g\left(\mu_i k_F^n,\mu_i k_F^p\right)\right\rbrace ,
\end{split}
\end{equation}
where $g(x_1,x_2) $ can be written as
\begin{equation}\label{gognyg}
\begin{split}
g\left(x_1,x_2\right)&=\frac{x_1^2-x_1x_2+x_2^2-2}{x_1^3+x_2^3}e^{-\frac{\left(x_1+x_2\right)^2}{4}}-2\frac{x_1^2+x_1x_2+x_2^2-2}{x_1^3+x_2^3}e^{-\frac{\left(x_1-x_2\right)^2}{4}}-\sqrt{\pi}\frac{x_1^3-x_2^3}{x_1^3+x_2^3}\text{erf}\left(\frac{x_1-x_2}{2}\right)\\
&+\sqrt{\pi}\text{erf}\left(\frac{x_1+x_2}{2}\right).\\
\end{split}
\end{equation}
In the GHF model, the symmetry energy is expressed as~\cite{Che12,Sel14}
\begin{equation}\label{gognyESYM}
\begin{split}
E_{\text{sym}}\left(\rho\right)&=\frac{\hbar^2k_F^2}{6m}-\frac{1}{8}t_0\left(1+2x_0\right)\rho^{1+\alpha}+\sum_{i=1,2}\Bigg\lbrace\frac{1}{72\sqrt{\pi}}\Big[\frac{4\cdot3^\frac{2}{3}}{\mu_ik_F}\left(2B_i-2H_i-4M_i+W_i\right)\left(e^{-k_F^2\mu_i^2}-1\right)\\
&+12k_F\mu_i\left(2B_i-H_i-2M_i+W_i\right)\left(e^{-k_F^2\mu_i^2}-1\right)-6k_F^3\mu_i^3\left(2H_i+W_i\right)\Big]\Bigg\rbrace ,
\end{split}
\end{equation}
and the fourth-order symmetry energy is expressed as
\begin{equation}\label{gognyEsym4}
\begin{split}
&E_{\text{sym,4}}\left(\rho\right)=\frac{\hbar^2k_F^2}{162m}+\frac{1}{324\sqrt{\pi}} \sum_{i=1,2}\Bigg\{ e^{-k_F^2\mu_i^2}\Big[\frac{14}{k_F\mu_i}\left(2B_i-2H_i-4M_i+W_i\right) -2k_F\mu_i\left(10H_i+20M_i-7W_i-14B_i\right)\\
&-\left(7k_F^3\mu_i^3+2k_F^5\mu_i^5\right)\left(H_i+2M_i-W_i-2B_i\right)\Big] -\Big[\frac{14}{k_F\mu_i}\left(2B_i-2H_i-4M_i+W_i\right) +\left(8k_F\mu_i-k_F^3\mu_i^3\right)\left(H_i+2M_i\right)\Big]\Bigg\} .
\end{split}
\end{equation}
\end{widetext}

\subsection{The MDI model}

In the MDI model, the interaction is the isospin- and momentum-dependent MDI interaction, which is a generalized isospin-dependent version of the momentum-dependent Yukawa interaction (MDYI)~\cite{Wel88}. The MDI interaction has been extensively applied in transport model simulations for heavy-ion collisions. The detail can be found in Refs.~\cite{Das03,LiBA04,Che05,Che14}. The EOS of asymmetric nuclear matter in the MDI model reads
\begin{equation}\label{MDIE}
E(\rho,\delta) = E_{k}(\rho,\delta) + \frac{V(\rho,\delta )}{\rho},
\end{equation}
where the kinetic energy contribution $E_{k}(\rho,\delta)$ can be obtained as
\begin{widetext}
\begin{eqnarray}
E_{k}(\rho,\delta) = \frac{1}{\rho}\int d^{3}{\vec{p}}
\left[\frac{p^{2}}{2
m}f_n(\vec{r},\vec{p}) + \frac{p^{2}}{2m} f_p(\vec{r},\vec{p})\right] %\notag \\
= \frac{4 \pi}{5 m h^3 \rho} (p^5_{F,n} + p^5_{F,p}),
\end{eqnarray}
and the potential energy density $V(\rho ,\delta )$ can be expressed as
\begin{eqnarray}
V(\rho ,\delta ) =\frac{A_{u}(x)\rho _{n}\rho _{p}}{\rho _{0}}+\frac{A_{l}(x)}{%
2\rho _{0}}(\rho _{n}^{2}+\rho _{p}^{2})+\frac{B}{\sigma +1}\frac{\rho
^{\sigma +1}}{\rho _{0}^{\sigma }}  %\notag \\
(1-x\delta ^{2})+\frac{1}{\rho _{0}}\sum_{\tau ,\tau ^{\prime
}}C_{\tau ,\tau ^{\prime }}  %\notag \\
\int \int d^{3}pd^{3}p^{\prime }\frac{f_{\tau }(\vec{r},\vec{p}%
)f_{\tau ^{\prime }}(\vec{r},\vec{p}^{\prime
})}{1+(\vec{p}-\vec{p}^{\prime })^{2}/\Lambda ^{2}}. \label{MDIV}
\end{eqnarray}%
Here $f_{\tau}(\vec{r},\vec{p})=\frac{2}{h^{3}}\Theta (p_{F,\tau}-p)$ is the nucleon phase space distribution function in nuclear matter at zero temperature  with $p_{F,\tau}=\hbar(3\pi^2\rho_\tau)^{1/3}$ being the Fermi momentum of nucleons of isospin $\tau$. $A_u(x)$, $A_l(x)$, $B$, $C_l$, $C_u$, $\Lambda$, $x$ are model parameters. The integral in Eq.~(\ref{MDIV}) can be obtained analytically at zero temperature~\cite{Che07}. In the MDI model, the symmetry energy can be expressed as~\cite{XuJ09,Che09}
\begin{equation}\label{MDIEsym}
\begin{split}
E_{\text{sym}}(\rho)&=\frac{8\pi}{9mh^3\rho}k_F^5+\frac{\rho}{4\rho_0}\left(A_l-A_u\right)-\frac{Bx}{\sigma+1}\left(\frac{\rho}{\rho_0}\right)^\sigma+\frac{C_l}{9\rho_0\rho}\left(\frac{4\pi}{h^3}\right)^2\Lambda^2\left[4k_F^4-\Lambda^2k_F^2\ln\frac{4k_F^2+\Lambda^2}{\Lambda^2}\right]\\
&+\frac{C_u}{9\rho_0\rho}\left(\frac{4\pi}{h^3}\right)^2\Lambda^2\Bigg[4k_F^4 -k_F^2\left(4k_F^2+\Lambda^2\right)\ln\frac{4k_F^2+\Lambda^2}{\Lambda^2}\Bigg],
\end{split}
\end{equation}
and the fourth-order symmetry energy can be written as~\cite{Che09}
\begin{equation}\label{MDIEsym4}
\begin{split}
E_{\text{sym,4}}(\rho)&=\frac{8\pi}{3^5mh^3\rho}k_F^5 -\frac{C_l}{3^5\rho_0\rho}\left(\frac{4\pi}{h^3}\right)^2 \Lambda^2\Bigg[7\Lambda^2k_F^2\ln\frac{4k_F^2+\Lambda^2}{\Lambda^2} -\frac{4\left(7\Lambda^4k_F^4+42\Lambda^2k_F^6+40k_F^8\right)}{\left(4k_F^2+\Lambda^2\right)^2}\Bigg]\\
&-\frac{C_u}{3^5\rho_0\rho}\left(\frac{4\pi}{h^3}\right)^2 \Lambda^2\Bigg[\left(7\Lambda^2k_F^2+16k_F^4\right) \ln\frac{4k_F^2+\Lambda^2}{\Lambda^2}-28k_F^4-\frac{8k_F^6}{\Lambda^2}\Bigg].
\end{split}
\end{equation}
\end{widetext}

\subsection{Determination of interaction parameters}
In the four mean-field models introduced above, one can determine their model parameters using some macroscopic quantities whose empirical values and uncertainties are available. This method has been successfully used to construct interaction parameter sets for nuclear energy density functionals and to study the correlations of
experimental observables with these macroscopic quantities~\cite{Che10,Zha13,Zha14PRC}. Note that since the spin-orbit coupling terms in the SHF, eSHF, GHF and MDI models are irrelevant for the properties of infinite nuclear matter, we omit them in the following.

In the SHF model, the $9$ interaction parameters  $ x_0$, $x_1$,$ x_2$, $x_3$, $t_0$, $t_1$, $t_2$, $t_3$ and $\alpha $ can be analytically expressed in terms of $9$ macroscopic quantities, namely, $ \rho_0$, $E_0(\rho_0)$, $K_0$, $E_{\text{sym}}(\rho_0)$, $L$, $m_{s,0}^*$, $m_{v,0}^*$, $G_S$ and $G_V$~\cite{Che10}.
Here $m_{s,0}^* = m_{s}^*(\rho_0)$ is the isoscalar nucleon effective mass at $\rho_0$, $m_{v,0}^* = m_{v}^*(\rho_0)$ is the isovector nucleon effective mass at $\rho_0$, $G_S$ is the gradient coefficient, and $G_V$ is the symmetry gradient coefficient~\cite{Che10}.
With these macroscopic parameters, we can evaluate the $E_{\text{sym}}(\rho)$, $E_{\text{sym,4}}(\rho)$ and other nuclear matter properties within the SHF model.

In the eSHF model, following Ref.~\cite{Zha16}, we set $\beta=1$ and $\gamma=1 $ and select $13$ macroscopic quantities $\rho_0$, $E_0(\rho_0)$, $K_0$, $J_0$, $E_{\text{sym}}(\rho_0)$, $L$, $K_{\rm {sym}}$, $m_{s,0}^*$, $m_{v,0}^*$, $G_S$, $G_V$, $G_{SV}$ and $G_0^{'}$~\cite{Zha16,Chen11} to determine the $13$ model parameters $x_0$, $x_1$, $x_2$, $x_3$, $x_4$, $x_5$, $t_0$, $t_1$, $t_2$, $t_3$, $t_4$, $t_5$ and $\alpha$. Here $G_{SV}$ is the cross gradient coefficient and $G_0^{'}$ is the Landau parameter~\cite{Zha16}.

In the GHF model, we fix the five Gaussian function parameters which are introduced to simulate the short-range nuclear force, namely, $W_1=-2047.16$ MeV, $B_1=-1700.00$ MeV, $H_1=-2414.93$ MeV, $M_1=1519.35$ MeV, and $\mu_1=0.8$ fm, and these values are taken from the parameter set D1N~\cite{Cha08}. We note that choosing the values of the five Gaussian function parameters from other parameter sets (e.g., D1S~\cite{Ber91}) does not change our conclusion. The other $8$ parameters, i.e., $W_2$, $B_2$, $H_2$, $M_2$, $\mu_2$, $t_0$, $\alpha$ and $x_0 $ can be determined explicitly in terms of $8$ macroscopic quantities, namely, $\rho_0$, $E_0(\rho_0)$, $K_0$, $U_{0,1k}$, $E_{\text{sym}}(\rho_0)$, $L$, $m_{s,0}^*$, and $m_{v,0}^*$, where $U_{0,1k}=U(\rho_0, E_K=1000~\text{MeV})$ is the single-nucleon potential at kinetic energy $1000$ MeV in SNM at $\rho_0$.

In the MDI model,
the $8$ macroscopic quantities $\rho_0$, $E_0(\rho_0)$, $K_0$, $U_{0,1k}$, $E_{\text{sym}}(\rho_0)$, $L$, $m_{s,0}^* $ and $m_{v,0}^*$ are used to determine the $8$ model parameters $A_u$, $ A_l$, $B$, $\sigma$, $C_u$, $C_l$, $\Lambda$, and $x$.

\begin{table}[tbp]
\caption{The empirical value (Emp.) with the corresponding uncertainty (in one standard deviation) for chosen macroscopic quantities and the linear-correlation coefficient $C_{AB}$ with $E_{\mathrm{sym,4}}(\rho_0)$ in the SHF model.}
\label{SHFMACRO}\centering
\begin{tabular}{*{6}{l l c c}}
\hline
\hline
$\qquad$	& 	Quantity  & Emp.  & $\quad$ $C_{AB}$\\
\hline
1 & $\rho_0(\text{fm}^{-3})$    & $0.16\pm 0.01$     &$\quad$-0.15	 \\
2 & $ E_0(\text{MeV}) $		& $-16.0\pm 1.0$	    &$\quad$ 0.00\\
3 & $ K_0(\text{MeV}) $		& $230.0\pm 25.0$ &$\quad$ 0.00	\\
4 & $ E_{\text{sym}}(\rho_0)(\text{MeV})  $		& $32.3\pm 1.0$ &$\quad$ 0.00	\\
5 & $L (\text{MeV})$   & $45.2\pm10.0$ &$\quad$ 0.00	\\
6 &  $ m^*_{s,0}/ \text{m} $		& $0.7\pm0.1$ &$\quad$ 0.41   \\
7 & $ m^*_{v,0}/ \text{m} $		& $0.6\pm0.1	$  &$\quad$ -0.85\\
8 & $G_S (\text{MeV} \cdot \text{fm}^5)$    & $132.0\pm30.0$ &$\quad$ 0.00	\\
9 & $ G_V(\text{MeV} \cdot \text{fm}^5) $		& $5.0\pm75.0$ &$\quad$ 0.01	\\

\hline
\end{tabular}
\end{table}

\begin{table}[tbp]
\caption{Similar with TABLE \ref{SHFMACRO} but in the eSHF model.}
\label{eSHFMACRO}\centering
\begin{tabular}{*{6}{l l c c}}
\hline
\hline
$\qquad$ & Quantity & Emp.   & $\quad$ $C_{AB}$ \\
\hline
1 & $\rho_0(\text{fm}^{-3})$    & $0.16\pm0.01$ &$\quad$ -0.05	 \\
2 & $ E_0(\text{MeV}) $		& $-16.0\pm1.0$  &$\quad$ 0.00  \\
3 & $ K_0(\text{MeV}) $		& $230.0\pm25.0$ &$\quad$ 0.00	\\
4 & $ J_0(\text{MeV}) $		& $-355.0\pm95.0$ &$\quad$ 0.01	\\
5 & $ E_{\text{sym}}(\rho_0)(\text{MeV})  $		& $32.3\pm1.0$ &$\quad$ 0.00	\\
6 & $L (\text{MeV})$   & $45.2\pm10.0$ &$\quad$ 0.00	\\
7 & $ K_{\text{sym}}(\text{MeV})  $		& $-100\pm165$ &$\quad$ -0.01	\\
8 & $ m^*_{s,0}/ \text{m} $		&$ 0.7\pm0.1	$ & $\quad$ 0.39   \\
9 & $ m^*_{v,0}/ \text{m} $		& $0.6\pm0.1$ &$\quad$ -0.82	 \\
10 & $G_S (\text{MeV} \cdot \text{fm}^5)$    & $132.0\pm30.0$ &$\quad$ 0.00	\\
11 & $ G_V(\text{MeV} \cdot \text{fm}^5) $		& $5.0\pm75.0$ & $\quad$ 0.00	\\
12 & $ G_{SV}(\text{MeV} \cdot \text{fm}^5) $		& $-8.0\pm20.0$ &$\quad$ 0.00	\\
13 & $ G_0^{'} $		& $0.8\pm0.8$ &$\quad$ 0.00	\\

\hline
\end{tabular}
\end{table}

\begin{table}[tbp]
\caption{Similar with TABLE \ref{SHFMACRO} but in the GHF model.}
\label{GOGNYMACRO}\centering
\begin{tabular}{*{6}{l l c c}}
\hline
\hline
$\qquad$   & Quantity   & Emp.  & $\quad$ $ C_{AB} $\\
\hline
1 & $ \rho_0 (\text{fm}^{-3})$		& $0.16\pm0.01$ &$\quad$ -0.07 \\
2 & $ E_0(\text{MeV}) $		& $-16.0\pm1.0$ &$\quad$ 0.00   \\
3 & $ K_0(\text{MeV}) $		& $230.0\pm25.0$ &$\quad$ 0.00	\\
4 & $ E_{\text{sym}}(\rho_0) (\text{MeV}) $		&$ 32.3\pm1.0$ &$\quad$ 0.00	\\
5 & $L(\text{MeV}) $    & $45.2\pm10.0$ &$\quad$ 0.00	\\
6 & $ m^*_{s,0}/ \text{m} $		& $0.7\pm0.1$ &$\quad$ 0.59    \\
7 & $ m^*_{v,0}/ \text{m} $		& $0.6\pm0.1$ &$\quad$ -0.74	 \\
8 & $ U_{0,1k}(\text{MeV}) $		& $55.0\pm10.0$ &$\quad$ 0.03	 \\

\hline
\end{tabular}
\end{table}

\begin{table}[tbp]
\caption{Similar with TABLE \ref{SHFMACRO} but in the MDI model.}
\label{MDIMACRO}\centering
\begin{tabular}{*{6}{l l c c}}
\hline
\hline
$\qquad$   & Quantity   & Emp.  & $\quad$ $C_{AB}$\\
\hline
1 & $ \rho_0 (\text{fm}^{-3})$		& $0.16\pm0.01$ &$\quad$ 0.01   \\
2 & $ E_0 (\text{MeV})$		& $-16.0\pm1.0$ &$\quad$ 0.00	    \\
3 & $ K_0(\text{MeV}) $		& $230.0\pm25.0$ &$\quad$ 0.00	\\
4 & $ E_{\text{sym}}(\rho_0) (\text{MeV}) $		& $32.3\pm1.0$ &$\quad$ 0.00	\\
5 & $L(\text{MeV})$    & $46.0\pm4.5$ & $\quad$ 0.00\\
6 & $ m^*_{s,0}/ \text{m} $		& $0.7\pm0.1	$ &$\quad$ 0.67    \\
7 & $ m^*_{v,0}/ \text{m} $		& $0.6\pm0.1$ &$\quad$ -0.70	 \\
8 & $ U_{0,1k}(\text{MeV}) $		& $55.0\pm10.0$ &$\quad$ 0.03	 \\

\hline
\end{tabular}
\end{table}

The empirical values together with their corresponding uncertainties
of the chosen macroscopic quantities for the SHF, eSHF, GHF and MDI models are shown in Tables~\ref{SHFMACRO}, \ref{eSHFMACRO}, \ref{GOGNYMACRO} and \ref{MDIMACRO}, respectively. It should be pointed out that the choose for the empirical values of the macroscopic quantities in the four tables is somewhat arbitrary but essentially reflect our current knowledge on these macroscopic quantities.
Insignificant variation of these empirical values does not change our present conclusion.
In particular,
we adopt the value $K_0=230.0\pm25.0$ MeV to be consistent with the constraints extracted from analyzing experimental data on giant monopole resonances of heavy nuclei~\cite{You99,Pie02,Agr03,Col04,Shl06,Gar07}.
The values $E_{\text{sym}}(\rho_0)=32.3\pm1.0$ MeV and $L=45.2\pm10.0$ MeV
are taken from the constraints obtained in Ref.~\cite{Zha13} by analyzing the isotope binding energy difference and neutron skin thickness,
while the values $m_{s,0}^*/m=0.7\pm0.1 $ and $m_{v,0}^*/m=0.6\pm0.1$
are chosen to be consistent with the extraction from global nucleon optical potentials constrained by world data on nucleon-nucleus and (p,n) charge-exchange reactions~\cite{XuC10,LiXH15} (see also Refs.~\cite{Jam89,LiChen15}).
The value $J_0 = -355.0\pm95.0$ MeV is taken from the estimate in Ref.~\cite{Chen11} by a correlation analysis method within SHF energy density functional, and the empirical values $G_S = 132.0 \pm 30.0~\text{MeV} \cdot \text{fm}^5$, $G_V = 5.0 \pm 75.0~\text{MeV} \cdot \text{fm}^5$, $G_{SV} = -8.0 \pm 20~\text{MeV} \cdot \text{fm}^5$ and $G_0^{'} = 0.8 \pm 0.8$ are taken from Refs.~\cite{Zha16,Che10}.
For the MDI and GHF models with momentum-dependent/finite range interactions,
the value $U_{0,1k} = 55.0\pm10.0$ MeV for the single-particle potential at kinetic energy $1000$ MeV in SNM at $\rho_0$ is taken to be in agreement with the nucleon optical potential extracted from the nucleon-nuclei scattering by Hama {\it at el.}~\cite{Ham90}.

We would like to emphasize that
here the uncertainty for the macroscopic quantities corresponds to one standard deviation (i.e., $1\sigma$), and as will be seen in the following, the value of each macroscopic quantity is actually obtained by Monte Carlo sampling according to a Gaussian distribution with the mean value and standard deviation equaling to its empirical value and uncertainty, respectively. That means the values of the macroscopic quantities can be significantly beyond the $1\sigma$ uncertain region of their empirical values. For example, the value $E_{\text{sym}}(\rho_0) = 32.3\pm1.0$ MeV means that for $99.7\%$ ($3\sigma$ uncertain region) samples of parameter sets, the value of $E_{\text{sym}}(\rho_0)$ is in the range of $29.3\sim35.3$ MeV.

\subsection{Statistical analysis}

In this work, we use the standard statistical analysis
method to determine the uncertainty of $E_{\mathrm{sym,4}}(\rho_0)$ and estimate its correlation with other macroscopic quantities.
Following Ref.~\cite{Bed15}, by assuming that the value of each macroscopic quantity follows a Gaussian distribution with the mean value and standard deviation equalling to, respectively, its empirical value and uncertainty as shown in Tables~\ref{SHFMACRO}, \ref{eSHFMACRO}, \ref{GOGNYMACRO} and \ref{MDIMACRO}, we sample randomly $0.1$ million parameter sets for each mean-field model using Monte Carlo method. For each sample of parameter set, the $E_{\mathrm{sym,4}}(\rho_0)$ can be evaluated, and thus we can obtain the histogram of the $E_{\mathrm{sym,4}}(\rho_0)$ distribution for each mean-field model using the corresponding $0.1$ million samples of parameter sets. From the histogram, we can obtain the mean value and standard deviation of the $E_{\mathrm{sym,4}}(\rho_0)$.

The linear-correlation coefficient $C_{AB}$ between the $E_{\text{sym,4}}(\rho_0) $ (A) and a macroscopic quantity (B) can be also estimated using the standard statistical method, i.e.,
\begin{eqnarray}
  C_{AB} &=& \frac{cov(A,B)}{\sigma(A)\sigma(B)},\label{CAB}\\
  cov(A,B) &=& \frac{1}{N-1}\sum_{i}(A_i- \langle A \rangle)(B_i- \langle B \rangle),\label{COV}\\
  \sigma(X)&=& \sqrt{\frac{1}{N-1}\sum_{i}(X_i- \langle X \rangle)^2},~(X=A, B)~~~~~\label{sigmaX}\\
  \langle X \rangle &=& \frac{1}{N}\sum_{i}X_i, (i=1, \cdot\cdot\cdot, N) \label{X}
\end{eqnarray}
where $cov(A,B)$ is the covariance between $A$ and $B$, $\sigma(X)$ is the standard deviation of $X$, $\langle X \rangle $ is the sample mean, and $N$ is the sample number. We would like to point out that the present estimate method of the linear-correlation coefficient $C_{AB}$ is different from the covariance analysis method (see, e.g., Ref.~\cite{Rei10}) where the quantities are constrained by some experimental data, but here the quantities are assumed to be independent random variables with their values following a Gaussian distribution.

\section{Results and discussions}\label{Sec3}

\begin{figure}[tbp]
\includegraphics[scale=0.34]{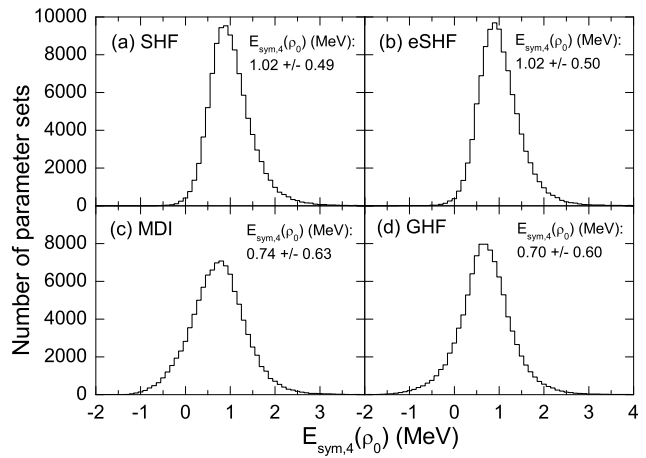}
\caption{Histogram of the number of parameter sets as a function of the value of $E_{\text{sym,4}}(\rho_0)$, from a sample of $0.1$ million parameter sets for each model. The average values of $E_{\text{sym,4}}(\rho_0)$ are also shown. }
\label{esym4rho0}
\end{figure}

Figure~\ref{esym4rho0} shows the histogram of the $E_{\text{sym,4}}(\rho_0)$ value with $0.1$ million samples of different interaction parameter sets given by Monte Carlo method, for each energy density functional, namely, the SHF model, the eSHF model, the GHF model and the MDI model. One sees that the histograms are very close to Gaussian distribution, and this is what we expect since the value of each macroscopic quantity we used to obtain the parameter sets is randomly independent and follows a Gaussian distribution. With these samples of interaction parameter sets, $E_{\text{sym,4}}(\rho_0)$ are estimated to be $1.02\pm0.49$ MeV, $1.02\pm0.50$ MeV, $0.70\pm0.60$ MeV and $0.74\pm0.63 $ MeV in the SHF, eSHF, GHF and MDI models, respectively.
Here the uncertainty of $E_{\text{sym,4}}(\rho_0)$ is the statistical standard deviation (i.e., 1-sigma) calculated by Eq. (\ref{sigmaX}).
One can see that the average values of $E_{\text{sym,4}}(\rho_0)$ in the MDI and GHF models are smaller than those in the SHF and eSHF models by about $0.3$ MeV. Our results indicate that the predicted $E_{\text{sym,4}}(\rho_0)$ in the four non-relativistic models are essentially less than $2$ MeV (within about 3-sigma), which is consistent with results from the relativistic mean-field model~\cite{Cai12} and the chiral pion-nucleon dynamics~\cite{Kai15}.

Using the $0.1$ million samples of different interaction parameter sets for each energy density functional, we also analyze the correlation between the $E_{\text{sym,4}}(\rho_0)$ and each macroscopic quantity.
The resulting correlation coefficients $C_{AB}$ in the four mean-field models are shown in Tabs.~\ref{SHFMACRO}, \ref{eSHFMACRO}, \ref{GOGNYMACRO} and \ref{MDIMACRO}, respectively.
It is interesting to see that the $E_{\text{sym},4}(\rho_0)$
is positively correlated to the $m_{s,0}^*$ with the
correlation coefficient $C_{AB}$ being $0.41$, $0.39$, $0.59$ and $0.67$ in the SHF, eSHF, GHF and MDI models, respectively, while negatively correlated to the $m_{v,0}^*$ with $C_{AB}$ equaling to $-0.85$, $-0.82$, $-0.74$ and $-0.70$, respectively. In addition, one can see a weak correlation between $E_{\text{sym},4}(\rho_0)$  and $\rho_0$. For the GHF and MDI models, a weak correlation is also found to exist between $E_{\text{sym},4}(\rho_0)$ and $U_{0,1k}$. Otherwise, there is essentially no correlation between the $E_{\text{sym},4}(\rho_0)$ and other macroscopic quantities for all models considered here.

Very interestingly, for the SHF and eSHF models, we find that the $E_{\text{sym},4}(\rho)$ can be simply expressed as
\begin{equation}\label{Esym4ofmsmv}
E_{\text{sym},4}(\rho)=\frac{\hbar^2}{162m}\left(\frac{3\pi^2\rho}{2}\right)^{\frac{2}{3}}\left[\frac{3m}{m_v^*(\rho)}-\frac{2m}{m_s^*(\rho)}\right].
\end{equation}
Using this formula, one can estimate the ratio of the correlation coefficient between $m_{s,0}^*$ and $E_{\text{sym},4}(\rho_0)$ to that between $m_{v,0}^*$ and $E_{\text{sym},4}(\rho_0)$ as $\frac{2}{m_{s,0}^{2}}/\frac{3}{m_{v,0}^{2}}\approx 1/2$, which is nicely consistent with the results in Tabs.~\ref{SHFMACRO} and \ref{eSHFMACRO}. In the GHF and MDI models, the situation is much more complicated, but the $E_{\text{sym},4}(\rho_0)$ also shows strong positive correlation with $m_{s,0}^*$ and strong negative correlation with $m_{v,0}^*$. In addition, from Eq.~(\ref{Esym4ofmsmv}), one can also easily understand why the values of $E_{\text{sym},4}(\rho_0)$ in the SHF and eSHF models are generally less than $2$ MeV. This is because a larger $E_{\text{sym},4}(\rho_0)$ generally needs a very small $m_{v,0}^*$ but large $m_{s,0}^*$, with which the models may fail to describe some experimental data (see, e.g., Ref.~\cite{Zha16mass}). Indeed, we note that the large values of the $E_{\text{sym},4}(\rho_0)$ shown in Fig.~\ref{esym4rho0} (a) and (b) are from the samples of parameter sets with very small $m_{v,0}^*$ but large $m_{s,0}^*$. Our results thus clearly demonstrate that the $E_{\text{sym},4}(\rho_0)$ is strongly correlated with $m_{v,0}^*$ and $m_{s,0}^*$.

\begin{figure}[tbp!]
\includegraphics[scale=0.34]{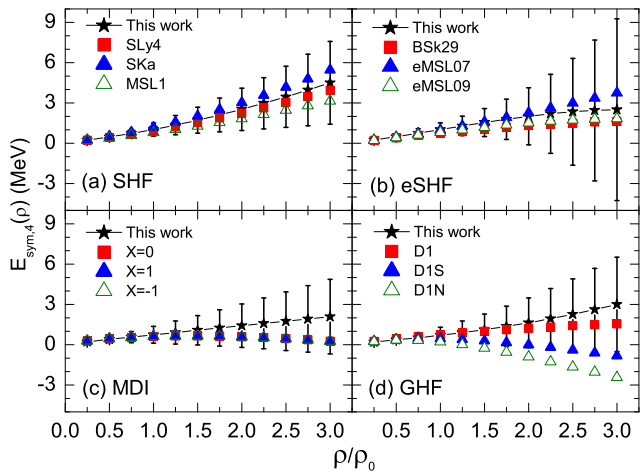}
\caption{(Color online) Density dependence of the fourth-order symmetry enrgy $ E_{\text{sym,4}}(\rho) $. The black stars with error bars are the results in this work. For comparison, the predictions of some typical interaction parameter sets in the literature,
i.e., SLy4~\cite{Cha98}, SKa~\cite{Koh76}, MSL1~\cite{Zha13}, BSk29~\cite{Gor15}, eMSL07~\cite{Zha16}, eMSL09~\cite{Zha16}, MDI with $x=1,0,-1$~\cite{Che05}, D1~\cite{Dec80}, D1S~\cite{Ber91}, and D1N~\cite{Cha08}, are also included for comparison.}
\label{esym4rho4}
\end{figure}

Similarly, the $E_{\text{sym,4}}(\rho) $ at other densities can also be obtained with the $0.1$ million samples of parameter sets for each model, and the results are displayed in Fig.~\ref{esym4rho4}.
One can see that the average value of the fourth-order symmetry energy always increases with increase of density and its density dependence is model dependent, especially at higher densities where the uncertainties are large. In particular, at three times saturation density, we obtain $E_{\text{sym,4}}(3\rho_0) =4.49\pm3.09$ MeV, $2.50\pm 6.76$ MeV, $2.99\pm3.52$ MeV and $2.09\pm2.78$ MeV in the SHF, eSHF, GHF and MDI models, respectively.

For comparison, we also show in Fig.~\ref{esym4rho4} the results from some typical interaction parameter sets, i.e.,
SLy4~\cite{Cha98}, SKa~\cite{Koh76} and MSL1~\cite{Zha13} in the SHF model; BSk29~\cite{Gor15}, eMSL07~\cite{Zha16} and eMSL09~\cite{Zha16} in the eSHF model; D1~\cite{Dec80}, D1S~\cite{Ber91} and D1N~\cite{Cha08} in the GHF model; and $x=0,1,-1$~\cite{Che05} in the MDI model.
It is seen that our prediction is consistent with essentially all the typical parameter sets considered here in 1-sigma uncertainty. However, the D1N predicts a much softer $E_{\text{sym},4}(\rho)$ at high densities than our estimate for the GHF model. This is due to the fact that the value of the single-nucleon potential $U_{0,1k}=-24.05$ MeV in D1N is much smaller than the value of $U_{0,1k}=55.0\pm10.0$ MeV extracted from nucleon-nuclei scattering by Hama {\it el at.}~\cite{Ham90} which is used in Tabs.~\ref{GOGNYMACRO} and \ref{MDIMACRO} for the GHF and MDI models. The different values of $U_{0,1k}$ make much difference in the results of $E_{\text{sym},4}(\rho)$ at high densities.

\section{Conclusion and outlook}\label{Sec4}

Within the framework of four energy density functionals of non-relativistic mean-field models, namely, the SHF, eSHF, GHF and MDI models, we have constructed large samples of the parameter sets by using Monte Carlo method and calculated the fourth-order symmetry energy with these samples of parameter sets. We have found that for these non-relativistic mean-field models, while the magnitude of the fourth-order symmetry energy are generally less than $2$ MeV at nuclear saturation density, its high-density behavior is model dependent and remains largely uncertain.

Furthermore, by analyzing the correlation between the $E_{\text{sym},4}(\rho_0)$ and other macroscopic quantities based on the samples of parameter sets, we have found that the $E_{\text{sym},4}(\rho_0)$ has a strong positive correlation with $m_{s,0}^*$ and a strong negative correlation with $m_{v,0}^*$. In particular, for the SHF and eSHF models, we have analytically expressed the $E_{\text{sym},4}(\rho)$ in terms of $m_{s}^*(\rho)$ and $m_{v}^*(\rho)$. Our results suggest that in the non-relativistic mean-field models, the $m_{v,0}^*$ and $m_{s,0}^*$ are the two key quantities to control the value of $E_{\text{sym},4}(\rho_0)$, and the small values of $E_{\text{sym},4}(\rho_0)$ observed in these models are mainly due to the larger $m_{v,0}^*$ values.

We would like to point out that one can also similarly estimate the value of other macroscopic quantities such as $K(\rho)$, $L(\rho)$, $E_{\text{sym}}(\rho)$ and so on, at other densities.
In addition, it will be extremely interesting to study the effects of nuclear short range correlations and tensor forces on the $E_{\text{sym,4}}(\rho)$ within the framework of beyond the mean-field approximation, and see whether a large $E_{\text{sym,4}}(\rho)$ is allowed or not. This will be very helpful to understand the properties of nuclear matter systems at extreme isospin, such as neutron stars.

\begin{acknowledgments}
This work was supported in part by the Major State Basic Research
Development Program (973 Program) in China under Contract Nos.
2013CB834405 and 2015CB856904, the National Natural Science
Foundation of China under Grant Nos. 11625521, 11275125 and
11135011, the Program for Professor of Special Appointment (Eastern
Scholar) at Shanghai Institutions of Higher Learning, Key Laboratory
for Particle Physics, Astrophysics and Cosmology, Ministry of
Education, China, and the Science and Technology Commission of
Shanghai Municipality (11DZ2260700).
\end{acknowledgments}

\end{document}